\documentclass[preprint,aps]{revtex4}%
\usepackage{amsfonts}
\usepackage{amsmath}
\usepackage{amssymb}
\usepackage{graphicx}%
\begin{document}
\preprint{ }
\title{Microscopic Theory of Spontaneous Decay in a Dielectric}
\author{Hao Fu, P. R. Berman}
\affiliation{Michigan Center for Theoretical Physics, FOCUS Center, and Physics Department,
University of Michigan, Ann Arbor, Michigan 48109-1040}

\begin{abstract}
The local field correction to the spontanous dacay rate of an impurity source
atom imbedded in a disordered dielectric is calculated to second order in the
dielectric density. The result is found to differ from predictions associated
with both "virtual" and "real" cavity models of this decay process. However,
if the contributions from two dielectric atoms at the same position are
included, the virtual cavity result is reproduced.

\end{abstract}
\volumeyear{ }
\volumenumber{ }
\issuenumber{ }
\eid{ }
\date{today}
\received[Received text]{date}

\revised[Revised text]{date}

\accepted[Accepted text]{date}

\maketitle

\section{Introduction}

The problem of spontaneous emission from an atom imbedded inside a dielectric
has attracted considerable interest \cite{int}. Most theoretical treatments of
this problem follow a macroscopic approach \cite{macro}. Based on different
models of the local environment of the imbedded atom, they give different
types of local field corrections to the spontaneous decay rate $\Gamma_{0}$ of
the impurity atom. The so-called virtual cavity model gives a decay rate,
$\Gamma_{virtual}=\sqrt{\epsilon}\left(  \frac{\epsilon+2}{3}\right)
^{2}\Gamma_{0}$, assuming that a virtual cavity surrounds the emitter, while
the "real" cavity model gives a rate, $\Gamma_{\operatorname{real}}%
=\sqrt{\epsilon}\left(  \frac{3\epsilon}{2\epsilon+1}\right)  ^{2}\Gamma_{0}$,
assuming that an empty spherical cavity surrounds the emitter. The quantity
$\epsilon$ is the permittivity of the dielectric, which is connected to the
microscopic polarizability $\alpha$ by the Lorentz-Lorenz relation
$\epsilon=1+\frac{N\alpha}{1-\frac{1}{3}N\alpha}$, where $N$ is the dielectric
density. Expansions for the decay rates in powers of $N\alpha$ yield
$\Gamma_{\operatorname{real}}=[1+\frac{7}{6}N\alpha+\frac{19}{72}(N\alpha
)^{2}+O(N\alpha)^{3}]\Gamma_{0}$ and $\Gamma_{virtual}=[1+\frac{7}{6}%
N\alpha+\frac{17}{24}(N\alpha)^{2}+O(N\alpha)^{3}]\Gamma_{0}$. To first order
in $N\alpha$, the real and virtual cavity models give identical results, but
they differ in higher order. To determine the validity range of these
macroscopic models, calculations using a somewhat more fundamental approach
are needed. Several attempts at such microscopic models involve (i) a
polariton approach for crystals \cite{polariton}, (ii) a Green's function
approach for crystals \cite{Green Crystal} and disordered dielectrics
\cite{Green Disorder}, and (iii) an amplitude approach for disordered
dielectrics \cite{Paul's}\cite{Crenshow's}. In the polariton method, the
interaction between the vacuum radiation field and the crystal atoms is solved
exactly; the eigenmodes of this system are the polaritons. The source atom
then decays by radiating polaritons. This polariton calculation agrees with
the virtual cavity result \cite{polariton}. In the Green's function approach,
the modification of the decay rate results from scattering of radiation
emitted from the source atom by the dielectric, calculated to all orders in
the dielectric density. This calculation reproduces the virtual cavity result
with the source atom at an interstitial position and the real cavity result
with source atom at a substitutional position in the crystal \cite{Green
Crystal}. For disordered dielectrics, the Green's function method gives the
virtual cavity result \cite{Green Disorder}. The amplitude method represents a
direct calculation of the modification of the decay rate as a perturbation
series in $N\alpha$ \cite{Paul's}\cite{Crenshow's}. To first order in
$N\alpha$, the radiation emitted by the source atom is scattered back to the
source atom by a single dielectric atom; the resultant decay rate agrees with
the both virtual and real cavity models to first order in $N\alpha$
\cite{Paul's}.

In this paper, the amplitude method is extended to second order by including
scattering events in which the radiation emitted by the source atom is
scattered back to the source atom by a combined scattering from two dielectric
atoms. It will be seen that the result differs from those of both the real and
virtual cavity models; however, when contributions to the decay rate
originating from scattering by two dielectric atoms located at the same
physical point are included, the calculation reverts to the virtual cavity model.

\section{Calculation of Second Order Contribution}

The source atom located at $\mathbf{R}=0$, has a $J=0$ ground state and a
$J=1$ excited state, the frequency separation of the ground and excited state
denoted by $\omega_{0}$. The uniformly distributed dielectric atoms have $J=0$
ground states and $J=1$ excited states, the frequency separation of the ground
and excited state denoted by $\omega.$ At $t=0$, the source atom is excited to
the $m=0$ excited state sublevel, the dielectric atoms are all in their ground
states, and there are no photons in the field. The process we consider is
radiation emitted by the source atom that is scattered by two dielectric atoms
back to the source atom. It is assumed that $\left\vert \omega-\omega
_{0}\right\vert /\Gamma_{0}\gg1$ but that $\left\vert \omega-\omega
_{0}\right\vert /\left(  \omega+\omega_{0}\right)  \ll1$ [rotating wave
approximation (RWA)].

We use a multipolar Hamiltonian \cite{Power}. The free part is
\begin{equation}
H_{0}=\frac{\hbar\omega_{0}}{2}\sigma_{z}+\sum_{j}\sum_{m=-1}^{1}\frac
{\hbar\omega}{2}\sigma_{z}^{(j)}(m)+\hbar\omega_{\mathbf{k}}a_{\mathbf{k}%
\lambda}^{\dagger}a_{\mathbf{k}\lambda}, \label{H0}%
\end{equation}
where $\sigma_{z}=\left(  \left\vert 2\right\rangle \left\langle 2\right\vert
-\left\vert 1\right\rangle \left\langle 1\right\vert \right)  $, $\left\vert
2\right\rangle $ and $\left\vert 1\right\rangle $ are the $m=0$ excited and
$J=0$ ground state eigenkets of the source atom, respectively, $\sigma
_{z}^{(j)}(m)=$ $\left(  \left\vert m\right\rangle ^{(j)}\left\langle
m\right\vert -\left\vert g\right\rangle ^{(j)}\left\langle g\right\vert
\right)  $ is the population difference operator between excited state
$\left\vert J=1,m\right\rangle $ and ground state $\left\vert
J=0,g\right\rangle $ \ of dielectric atom $j$, and $a_{\mathbf{k}\lambda}$ is
the annihilation operator for a photon having momentum $\mathbf{k}$ and
polarization $\lambda$. A summation convention is used, in which any repeated
symbol on the right hand side of an equation is summed over, unless it also
appears on the left-hand side of the equations.

The interaction part of the Hamiltonian is,%
\begin{equation}
V=-\mathbf{d}_{0}\cdot\frac{\mathbf{D(}0\mathbf{)}}{\epsilon_{0}}%
-\mathbf{d}_{j}\cdot\frac{\mathbf{D(R}_{j}\mathbf{)}}{\epsilon_{0}},
\label{Inter}%
\end{equation}
where $\mathbf{d}_{0}\ $and $\mathbf{d}_{j}$ are the dipole operators of the
source atom located at the origin and a dielectric atom located at position
$\mathbf{R}_{j}$ respectively. The operator $\mathbf{D}$ is the displacement
field having positive frequency component
\begin{equation}
\mathbf{D}_{+}\mathbf{(R)=}i\epsilon_{0}\sum_{\mathbf{k},\lambda}\sqrt
{\frac{\hbar\omega_{\mathbf{k}}}{2\epsilon_{0}V}}\epsilon_{\mathbf{k}%
}^{(\lambda)}a_{\mathbf{k}\lambda}e^{i\mathbf{k\cdot R}} \label{Field}%
\end{equation}
where $V$ is the quantization volume and $\epsilon_{\mathbf{k}}^{(\lambda)}$
is a unit polarization vector, with%
\begin{align}
\epsilon_{\mathbf{k}}^{(1)}  &  =\cos\theta_{\mathbf{k}}\cos\phi_{\mathbf{k}%
}\mathbf{\hat{x}}+\cos\theta_{\mathbf{k}}\sin\phi_{\mathbf{k}}\mathbf{\hat{y}%
}-\sin\theta_{\mathbf{k}}\mathbf{\hat{z}}\label{polarization}\\
\epsilon_{\mathbf{k}}^{(2)}  &  =-\sin\phi_{\mathbf{k}}\mathbf{\hat{x}}%
+\cos\phi_{\mathbf{k}}\mathbf{\hat{y}.} \label{Polarization}%
\end{align}
In the RWA, one can write%
\begin{equation}
V=\sum_{\mathbf{k}}\hbar g_{\mathbf{k}}(\sigma_{+}a_{\mathbf{k}}%
-a_{\mathbf{k}}^{\dagger}\sigma_{-})+\sum_{\mathbf{k},\lambda,m}%
\hbar(g_{\mathbf{k\lambda}}^{\prime}(m)\sigma_{+}^{(j)}(m)a_{\mathbf{k}%
\lambda}e^{i\mathbf{k\cdot R}}+g_{\mathbf{k\lambda}}^{\prime}(m)^{\ast
}a_{\mathbf{k}\lambda}^{\dagger}\sigma_{-}^{(j)}(m)e^{-i\mathbf{k\cdot R}})
\label{interaction}%
\end{equation}%
\begin{align}
g_{\mathbf{k}}  &  =-i\sqrt{\frac{\omega_{\mathbf{k}}}{2\hbar\epsilon_{0}V}%
}\mu(\epsilon_{\mathbf{k}}^{(\lambda)})_{0}\label{coupling}\\
g_{\mathbf{k\lambda}}^{\prime}  &  =-i\sqrt{\frac{\omega_{\mathbf{k}}}%
{2\hbar\epsilon_{0}V}}\mu^{\prime}(\epsilon_{\mathbf{k}}^{(\lambda)}%
)_{m}^{\ast},
\end{align}
where the $\sigma_{\pm}$ are raising and lowering operators for the source
atom and $\sigma_{\pm}^{(j)}(m)$ are raising and lowering operators between
the excited state $\left\vert J=1,m\right\rangle $ and the ground state
$\left\vert J=0,g\right\rangle $ of dielectric atom $j$, $\mu$ is the reduced
matrix element of the dipole operator $\mathbf{d}_{0}$ and $\mu^{\prime}$ is
that of $\mathbf{d}_{j}$ between ground and excited state manifolds. The
$(\epsilon_{\mathbf{k}}^{(\lambda)})_{\pm1}=\mp\frac{(\epsilon_{\mathbf{k}%
}^{(\lambda)})_{x}\pm i(\epsilon_{\mathbf{k}}^{(\lambda)})_{y}}{\sqrt{2}}$,
$(\epsilon_{\mathbf{k}}^{(\lambda)})_{0}=(\epsilon_{\mathbf{k}}^{(\lambda
)})_{z}$ are spherical components of the polarization vectors. The source atom
interacts only with the $z$ component of the radiation field.

The calculation proceeds as in Ref. \cite{Paul's}, with the addition of terms
that couple dielectric atoms to dielectric atoms via the radiation field.
After eliminating intermediate states involving the radiation field, one
arrives at%
\begin{subequations}
\begin{align}
\dot{b}_{2}  &  =-\gamma b_{2}-\gamma(\frac{\mu^{\prime}}{\mu})e^{i\Delta
t}G_{0,m_{j}}(\mathbf{R}_{j},\omega)b_{m_{j}}(t)\label{motiona}\\
\dot{b}_{m_{j}}  &  =-\gamma^{\prime}b_{m_{j}}-\gamma(\frac{\mu^{\prime}}{\mu
})e^{-i\Delta t}G_{m,0}(\mathbf{R}_{j},\omega_{0})b_{2}(t)-\gamma^{\prime
}G_{m_{j},m_{s}^{\prime}}(\mathbf{R}_{j}-\mathbf{R}_{s},\omega)b_{m_{s}%
^{\prime}}(t), \label{motionb}%
\end{align}
where $\gamma=2\mu^{2}\omega_{0}^{3}/3\hbar c^{3}$ and $\gamma^{\prime}%
=2\mu^{\prime2}\omega_{0}^{3}/3\hbar c^{3}$ are (half) the excited state decay
rate of the source and dielectric atoms, respectively, $b_{2}$ is the state
amplitude for the source atom to be in state $\left\vert 2\right\rangle
=\left\vert J=1,m=0\right\rangle $ and all dielectric atoms in their ground
states, and $b_{m_{j},}$ is the state amplitude for dielectric atom $j$ to be
in excited state $\left\vert J=1,m\right\rangle $ all other atoms in their
ground states. We have set $b_{2}(t-\tau)\approx b_{2}(t)$ and $b_{m_{j}%
}(t-\tau)\approx b_{m_{j}}(t)$ on the assumption that $\gamma R_{0}%
/c,\gamma^{\prime}R_{0}/c\ll1$, where $R_{0}$ is the sample size. The quantity
$G_{m_{j},m_{s}^{\prime}}(\mathbf{R}_{j}-\mathbf{R}_{s},\omega)$ is a
propagator for scattering from a dielectric atom in sublevel $m_{j}$ at
position $\mathbf{R}_{j}$ to one in sublevel $m_{s}$ at position
$\mathbf{R}_{s}$ given by%
\end{subequations}
\begin{equation}
G_{m_{j},m_{s}^{\prime}}(\mathbf{R,}\omega)=\frac{3}{8\pi}\frac{1}{\pi
\omega^{3}}\int_{0}^{t}d\tau\int_{-\infty}^{\infty}d\omega_{k}\omega_{k}%
^{3}e^{-i(\omega_{k}-\omega)\tau}\int d\Omega_{\mathbf{k}}(\epsilon
_{\mathbf{k}}^{(\lambda)})_{m_{j}}^{\ast}(\epsilon_{\mathbf{k}}^{(\lambda
)})_{m_{s}^{\prime}}e^{i\mathbf{k}\cdot\mathbf{R}}, \label{propagator}%
\end{equation}
while $G_{m,0}(\mathbf{R}_{j},\omega)$ is a propagator for scattering from the
source atom to a dielectric atom in sublevel $m_{j}$ at position
$\mathbf{R}_{j}$. In what follows we ignore the difference between $\omega
_{0}$ and $\omega$, consistent with the RWA.

In order to solve Eqs. (\ref{motiona},\ref{motionb}), we assume that $b_{2}$
varies slowly on the time scale $1/\Delta$. If $b_{m_{j}}=y_{m_{j}}e^{-i\Delta
t}$, Eqs.(\ref{motiona},\ref{motionb}) are transformed to
\begin{subequations}
\label{eqs}%
\begin{align}
\dot{b}_{2}  &  =-\gamma b_{2}-\gamma(\frac{\mu^{\prime}}{\mu})G_{0,m}%
(\mathbf{R}_{j},\omega_{0})y_{m_{j}}(t)\label{eqsa}\\
(\gamma^{\prime}-i\Delta)y_{m_{j}}  &  =-\gamma(\frac{\mu^{\prime}}{\mu
})G_{m,0}(\mathbf{R}_{j},\omega_{0})b_{2}(t)-\gamma^{\prime}G_{m_{j}%
,m_{s}^{\prime}}(\mathbf{R}_{j}-\mathbf{R}_{s},\omega_{0})y_{m_{s}^{\prime}%
}(t) \label{eqsb}%
\end{align}
The formal solution for $\dot{b}_{2}$ is
\end{subequations}
\begin{equation}
\dot{b}_{2}=-\gamma b_{2}+\gamma(\frac{\mu^{\prime}}{\mu})G_{0,m_{j}%
}(\mathbf{R}_{j},\omega_{0})\left[  \frac{1}{\gamma^{\prime}-i\Delta
+\gamma^{\prime}\mathbf{G}}\right]  _{m_{j},m_{s}^{\prime}}\gamma(\frac
{\mu^{\prime}}{\mu})G_{m_{s}^{\prime},0}(\mathbf{R}_{s},\omega_{0})b_{2}
\label{Infinite order}%
\end{equation}
where $\mathbf{G}$ is an $3N\times3N$ matrix having matrix elements
$G_{m_{j},m_{s}^{\prime}}(\mathbf{R}_{j}-\mathbf{R}_{s},\omega_{0})$. This can
be expanded as a power series in $N\alpha$ with $\alpha=-\frac{4\pi\mu
^{\prime2}}{\hbar\Delta}.$ To second order, one finds%
\begin{equation}
\dot{b}_{2,0}=-\gamma b_{2,0}\left[
\begin{array}
[c]{c}%
1+iN\alpha\frac{k_{0}^{3}}{6\pi N}G_{0,m_{j}}(\mathbf{R}_{j},\omega
_{0})G_{m_{j},0}(\mathbf{R}_{j},\omega_{0})\\
-(N\alpha)^{2}(\frac{k_{0}^{3}}{6\pi N})^{2}G_{0,m_{j}}(\mathbf{R}_{j}%
,\omega_{0})G_{m_{j},m_{s}^{\prime}}(\mathbf{R}_{j}-\mathbf{R}_{s},\omega
_{0})G_{m_{s}^{\prime},0}(\mathbf{R}_{s},\omega_{0})
\end{array}
\right]  \label{second order}%
\end{equation}
The term linear in the density gives the first order local field correction
$\frac{\delta\gamma^{(1)}}{\gamma}=\frac{7}{6}N\alpha$ \cite{Paul's}, which
agrees with both the virtual and real cavity models to this order.

We now calculate the second order correction,
\begin{equation}
\frac{\delta\gamma^{(2)}}{\gamma}=-(N\alpha)^{2}(\frac{k_{0}^{3}}{6\pi N}%
)^{2}G_{0,m_{j}}(\mathbf{R}_{j},\omega_{0})G_{m_{j},m_{s}^{\prime}}%
(\mathbf{R}_{j}-\mathbf{R}_{s},\omega_{0})G_{m_{s}^{\prime},0}(\mathbf{R}%
_{s},\omega_{0}).
\end{equation}
The sum over $\mathbf{R}_{j}$ and $\mathbf{R}_{s}$ can be converted to
integrals using $\sum\rightarrow N\int d\mathbf{R}$. In this manner, one finds%
\begin{equation}
\frac{\delta\gamma^{(2)}}{\gamma}=-(N\alpha)^{2}(\frac{k_{0}^{3}}{6\pi}%
)^{2}\int\int d\mathbf{R}_{2}d\mathbf{R}_{1}G_{0,m_{j}}(\mathbf{R}_{2}%
,\omega_{0})G_{m_{j},m_{s}^{\prime}}(\mathbf{R},\omega_{0})G_{m_{s}^{\prime
},0}(\mathbf{R}_{1},\omega_{0}) \label{sec}%
\end{equation}
where $\mathbf{R=}$\textbf{$R$}$_{2}-$\textbf{$R$}$_{1}.$ The next step is to
evaluate the $G_{m_{j},m_{s}^{\prime}}(\mathbf{R,}\omega)$. The details of the
calculation are given in the Appendix and one obtains
\begin{subequations}
\begin{align}
G_{11}  &  =\sqrt{4\pi}h_{0}(k_{0}R)Y_{0,0}(\mathbf{\hat{R}})-\frac{1}{2}%
\sqrt{\frac{4\pi}{5}}h_{2}(k_{0}R)Y_{2,0}(\mathbf{\hat{R}});\label{11}\\
G_{00}  &  =\sqrt{4\pi}h_{0}(k_{0}R)Y_{0,0}(\mathbf{\hat{R}})+\sqrt{\frac
{4\pi}{5}}h_{2}(k_{0}R)Y_{2,0}(\mathbf{\hat{R}});\\
G_{1,-1}  &  =-\frac{3}{2}\sqrt{\frac{8\pi}{15}}h_{2}(k_{0}R)Y_{2,-2}%
(\mathbf{\hat{R}});\\
G_{-1,1}  &  =-\frac{3}{2}\sqrt{\frac{8\pi}{15}}h_{2}(k_{0}R)Y_{2,2}%
(\mathbf{\hat{R}});\\
G_{1,0}  &  =-\frac{3}{2}\sqrt{\frac{4\pi}{15}}h_{2}(k_{0}R)Y_{2,-1}%
(\mathbf{\hat{R}});\label{10}\\
G_{-1,0}  &  =-\frac{3}{2}\sqrt{\frac{4\pi}{15}}h_{2}(k_{0}R)Y_{2,1}%
(\mathbf{\hat{R}}),
\end{align}
where $Y_{\ell,m}(\mathbf{\hat{R}})$ is a spherical harmonic and $k_{0}%
=\omega_{0}/c$. The remaining $G_{m_{j},m_{s}^{\prime}}$s are obtained using
$G_{-1,-1}=G_{11}$, $G_{0,-1}=-G_{1,0}$, and $G_{0,1}=-G_{-1,0}$ $.$ The
spherical Hankel functions of the first kind, $h_{0}(k_{0}R)$ and $h_{2}%
(k_{0}R),$ conform to the appropriate boundary conditions in which only
outgoing scattered waves are considered.

The calculation for \bigskip$\frac{\delta\gamma^{(2)}}{\gamma}$ is tedious,
since it involves contributions from nine terms. We will show how to calculate
one specific contribution, $m_{j}=1$, $m_{s}^{\prime}=1$, and then give the
final results for the other components. Substituting Eqs. (\ref{11},\ref{10})
in Eq. (\ref{sec}), we find
\end{subequations}
\begin{align}
\frac{\delta\gamma^{(2)}(1,1)}{\gamma} &  =(N\alpha)^{2}(\frac{k_{0}^{3}}%
{6\pi})^{2}\frac{6\pi^{\frac{3}{2}}}{5}\int\int d\mathbf{R}_{1}d\mathbf{R}%
_{2}h_{2}(k_{0}R_{2})Y_{2,1}(\mathbf{\hat{R}}_{2})h_{0}(k_{0}R_{21}%
)Y_{0,0}(\mathbf{\hat{R}}_{21})h_{2}(k_{0}R_{1})Y_{2,-1}(\mathbf{\hat{R}}%
_{1})\nonumber\\
&  -(N\alpha)^{2}(\frac{k_{0}^{3}}{6\pi})^{2}\frac{3\pi^{\frac{3}{2}}}%
{5\sqrt{5}}\int\int d\mathbf{R}_{1}d\mathbf{R}_{2}h_{2}(k_{0}R_{2}%
)Y_{2,1}(\mathbf{\hat{R}}_{2})h_{2}(k_{0}R_{21})Y_{2,0}(\mathbf{\hat{R}}%
_{21})h_{2}(k_{0}R_{1})Y_{2,-1}(\mathbf{\hat{R}}_{1}).\label{second order 11}%
\end{align}
To evaluate this, we expand the Hankel functions as \cite{Dano's}%
\begin{align}
h_{l}(k_{0}R_{21})Y_{l,m}(\mathbf{\hat{R}}_{21}) &  =i^{l_{1}+l_{2}%
-l}(-1)^{l_{2}+m}\sqrt{4\pi(2l+1)(2l_{1}+1)(2l_{2}+1)}\nonumber\\
&  \times\left[  \Theta\left(  R_{2}-R_{1}\right)  +\left(  -1\right)
^{l}\Theta\left(  R_{1}-R_{2}\right)  \right]  \left(
\begin{array}
[c]{ccc}%
l_{1} & l & l_{2}\\
0 & 0 & 0
\end{array}
\right)  \left(
\begin{array}
[c]{ccc}%
l_{1} & l & l_{2}\\
m_{1} & m & m_{2}%
\end{array}
\right)  \nonumber\\
\times &  h_{l_{1}}(k_{0}R_{>})j_{l_{2}}(k_{0}R_{<})Y_{\ell_{1},m_{1}%
}(\mathbf{\hat{R}}_{>})Y_{\ell_{2},m_{2}}(\mathbf{\hat{R}}_{<}%
)\label{hankel expansion1}%
\end{align}
where $\binom{...}{...}$ is a 3-$j$ symbol, $j_{l}(x)$ is a spherical Bessel
function, and $R_{>}$ $(R_{<})$ is the larger (smaller) of $R_{1}$ and $R_{2}%
$. When this expansion is used in Eq. (\ref{second order 11}), the angular
integration selects out only $l=2,0$ and $m=1,-1,0$ terms, such that
\begin{equation}
\frac{\delta\gamma^{(2)}(1,1)}{\gamma}=-\frac{(N\alpha)^{2}}{7}\int
_{0}^{\infty}d\mathbf{\rho}_{2}\mathbf{\rho}_{2}^{2}\int_{0}^{\rho_{2}%
}d\mathbf{\rho}_{1}\mathbf{\rho}_{1}^{2}h_{2}(\mathbf{\rho}_{2})h_{2}%
(\mathbf{\rho}_{1})h_{2}(\mathbf{\rho}_{2})j_{2}(\mathbf{\rho}_{1}%
)\label{correction for 11}%
\end{equation}
with $\mathbf{\rho}_{1}=k_{0}R_{1}$, $\mathbf{\rho}_{2}=k_{0}R_{2}$. To
evaluate the above integral, we add a convergence factor $e^{-\epsilon
\mathbf{\rho}_{2}}$, and eventually take the limit $\epsilon\rightarrow0$. The
imaginary part of the integral diverges as $\mathbf{\rho}_{2}\rightarrow0$,
but the real part is finite and gives the local field correction to the decay
rate. The result is
\[
\frac{\delta\gamma^{(2)}(1,1)}{\gamma}=\frac{15}{112}(N\alpha)^{2}%
\]
and the corresponding results for the other terms are \cite{note}
\begin{align*}
\frac{\delta\gamma^{(2)}(-1,-1)}{\gamma} &  =\frac{\delta\gamma^{(2)}%
(1,1)}{\gamma}=\frac{15}{112}(N\alpha)^{2},\\
\frac{\delta\gamma^{(2)}(0,0)}{\gamma} &  =\frac{25}{63}(N\alpha)^{2},\\
\frac{\delta\gamma^{(2)}(0,1)}{\gamma} &  =\frac{\delta\gamma^{(2)}%
(1,0)}{\gamma}=\frac{\delta\gamma^{(2)}(-1,0)}{\gamma}=\frac{\delta
\gamma^{(2)}(0,-1)}{\gamma}=\frac{3}{28}(N\alpha)^{2},\\
\frac{\delta\gamma^{(2)}(-1,1)}{\gamma} &  =\frac{\delta\gamma^{(2)}%
(1,-1)}{\gamma}=-\frac{3}{56}(N\alpha)^{2}.
\end{align*}
The total second order correction to the decay rate is
\begin{equation}
\frac{\delta\gamma^{(2)}}{\gamma}=2\frac{\delta\gamma^{(2)}(1,1)}{\gamma
}+\frac{\delta\gamma^{(2)}(0,0)}{\gamma}+4\frac{\delta\gamma^{(2)}%
(0,1)}{\gamma}+2\frac{\delta\gamma^{(2)}(-1,1)}{\gamma}=\frac{71}{72}%
(N\alpha)^{2}.\label{finres}%
\end{equation}
This result differs from both the virtual $\left[  \frac{51}{72}(N\alpha
)^{2}\right]  $ and real $\left[  \frac{19}{72}(N\alpha)^{2}\right]  $ cavity
models .

Our result can be compared with Fleischhauer's \cite{Green Disorder}$.$ The
Fourier transform of $G_{m,m^{\prime}}(\mathbf{R},\omega_{0})$ is the tensor
field propagator $\mathbf{F}^{(0)}(\mathbf{q},\omega)$ in his paper, differing
only by prefactors. The integral (\ref{sec}) can then be done in either
coordinate or momentum space. The momentum space integration gives a different
result than our coordinate space calculation above. This surprising
discrepancy can be explained by the way we expand $h_{l}(k_{0}R_{21}%
)Y_{l,m}(\mathbf{\hat{R}}_{21})$. The expansion we used is valid for
$R_{1}>R_{2}$ or $R_{2}>R_{1}$, but is not defined for $R_{1}=R_{2}$. For a
well-behaved integral this will not make any difference since $R_{1}=R_{2}$
contributes a set of measure zero. In the present case, however, where the
dipole-dipole interaction between dielectric atoms diverges when one atom is
on the top the other, i.e. when $\mathbf{R}_{1}=\mathbf{R}_{2}$, the
contribution from $R_{1}=R_{2}$ can be finite.

It is not easy to estimate this contribution in the original form of the
integral (\ref{second order 11}). Instead, it proves useful to Fourier
transform just \textit{one} of the $G_{m,m^{\prime}}$ in the integrand. As an
example, we consider the integral in the second term of Eq.
(\ref{second order 11})%
\begin{equation}
I=\int\int d\mathbf{R}_{2}d\mathbf{R}_{1}h_{2}(kR_{2})Y_{2,1}(\mathbf{\hat{R}%
}_{2})h_{2}(kR_{21})Y_{2,0}(\mathbf{\hat{R}}_{21})h_{2}(kR_{1})Y_{2,-1}%
(\mathbf{\hat{R}}_{1}) \label{11 integral}%
\end{equation}
We Fourier transform $h_{2}(kR_{21})Y_{2,0}(\mathbf{\hat{R}}_{21})e^{-\epsilon
R_{21}}$, using a convergence factor $e^{-\epsilon R_{21}}$ that is physically
connected with the boundary condition of outgoing spherical waves. Carrying
out the Fourier transform in Eq. (\ref{11 integral}), we find%
\begin{equation}
I=-\frac{4\pi i}{k^{3}}\int\int d\mathbf{R}_{2}d\mathbf{R}_{1}h_{2}%
(kR_{2})Y_{2,1}(\mathbf{\hat{R}}_{2})h_{2}(kR_{1})Y_{2,-1}(\mathbf{\hat{R}%
}_{1})\int\frac{d\mathbf{p}}{(2\pi)^{3}}\frac{p^{2}}{k^{2}-p^{2}+i\epsilon
}Y_{2,0}(\mathbf{\hat{p}})e^{i\mathbf{p\cdot(R}_{2}-\mathbf{R}_{1})}
\label{Partial Fourier}%
\end{equation}
The angular integrations can be done by expanding $e^{i\mathbf{p\cdot R}_{2}}%
$, $e^{-i\mathbf{p\cdot R}_{1}}$ in terms of spherical harmonics and Bessel
functions. In this manner one obtains%
\begin{equation}
I=\frac{1}{14}\sqrt{\frac{5}{\pi}}\frac{(4\pi)^{3}i}{k^{3}}\int\int
dR_{2}dR_{1}R_{2}^{2}R_{1}^{2}h_{2}(kR_{2})h_{2}(kR_{1})\int\frac{dp}%
{(2\pi)^{3}}\frac{p^{4}}{k^{2}-p^{2}+i\epsilon}%
\end{equation}
We are interested only in the contribution in the region where $R_{1}%
=R_{2\text{ }}$. This contribution can be isolated by integrating $R_{2}$ from
$R_{1}-a$ to $R_{1}+a$, and then integrating the resultant expression over $p$
using the method of residues. In the limit that both $a$ and $\epsilon$ tend
to zero, one obtains the contribution $\delta I$ from the region
$R_{1}=R_{2\text{ }}$ as
\begin{equation}
\operatorname{Re}[\delta I]=\operatorname{Re}[\frac{-1}{7}\sqrt{5\pi}\frac
{2i}{k^{3}}\int dR_{2}R_{2}^{2}h_{2}(kR_{2})h_{2}(kR_{2})]=\frac{-5\sqrt{5\pi
}}{7k^{6}} \label{deta}%
\end{equation}
(the imaginary part of $\delta I$ diverges). The contribution from the sphere
$R_{1}=R_{2\text{ }}$ is identical to that from $\mathbf{R}_{1}=\mathbf{R}%
_{2}$ since all other points with $\mathbf{R}_{1}\neq\mathbf{R}_{2\text{ }}$
on the sphere are regular and contribute zero to the integral. The same
calculation can be done for the first integral in the Eq.
(\ref{second order 11}). For this term, there is no contribution from the
region $R_{1}=R_{2\text{ }}$ (no delta function like term is found) since
$h_{0}(kR_{21})$ has a lower order divergence at $R_{21}=0$ than does
$h_{2}(kR_{21})$.

Including contributions of the type (\ref{deta}), we find
\begin{align}
\frac{\delta\gamma^{(2)}(0,0)}{\gamma}  &  =\frac{1}{3}(N\alpha)^{2}%
,\nonumber\\
\frac{\delta\gamma^{(2)}(1,1)}{\gamma}  &  =\frac{\delta\gamma^{(2)}%
(1,-1)}{\gamma}=\frac{7}{48}(N\alpha)^{2},\nonumber\\
\frac{\delta\gamma^{(2)}(0,1)}{\gamma}  &  =\frac{\delta\gamma^{(2)}%
(1,0)}{\gamma}=\frac{\delta\gamma^{(2)}(-1,0)}{\gamma}=\frac{\delta
\gamma^{(2)}(0,-1)}{\gamma}=\frac{1}{12}(N\alpha)^{2},\nonumber\\
\frac{\delta\gamma^{(2)}(-1,1)}{\gamma}  &  =\frac{\delta\gamma^{(2)}%
(1,-1)}{\gamma}=-\frac{1}{8}(N\alpha)^{2}. \label{other3}%
\end{align}
When these are summed the total $\frac{\delta\gamma^{(2)}}{\gamma}=\frac
{17}{24}(N\alpha)^{2}$ agrees with the virtual cavity result.

\section{Discussion}

The second order contribution to the modified spontaneous emission rate of an
impurity atom in a disordered dielectric has been calculated using a
microscopic theory. Depending on the manner in which overlapping atoms are
treated, one arrives at different results. If the delta function contributions
at $\mathbf{R}_{1}=\mathbf{R}_{2}$ are included, the virtual cavity model is
recovered, but if such terms are excluded, neither the real nor virtual cavity
model results are found. It seems to us somewhat of an open question at this
point as to whether or not such contributions can be uniquely calculated once
Eq. (\ref{Infinite order}) is expanded in a power series in the density. The
reason for this is that the expansion parameter is \textit{not }small as
interatomic distances tend to zero. That the expansion can lead to divergences
is already evident if the integrations are carried out using a different set
of variables \cite{note}. From physical considerations, however, the decay
rate does not diverge, even for interparticle spacings much less than a
wavelength. Actually, dielectric atoms within a sphere of radius
$\lambda\left(  \gamma^{\prime}/\Delta\right)  ^{1/3}$ reradiate collectively;
outside this radius, there is destructive interference resulting in some
additional finite contribution to the decay rate. In dealing with a
homogeneous dielectric, we have performed the ensemble average by integrating
over all space assuming a constant density. This averaging process includes
configurations where interparticle spacings are sufficiently small to
invalidate the expansion (\ref{second order}). Nevertheless, the procedure has
yielded finite results for the change in the decay parameter.

Different experiments support both the real and virtual cavity results
\cite{expt}. The source atom in these experiments is usually an impurity ion
in a protective molecular cage. No experiments of this nature have been
carried out with impurity atomic radiators in a dielectric that consists of a
dense atomic vapor. It may be possible to use an alkali metal atom as the
source atom and rare gas atoms as the dielectric atoms. With such a system,
one could not make the rotating wave approximation used in this paper, but the
physics is not changed in any substantive manner. The key feature of the
alkali metal - rare gas system is the extremely small quenching cross sections
for rare gas collisions to inelastically change the electronic state of the
alkali atom \cite{cross}. Any quenching cross sections would appear as a
modification of the decay rate that would mask the sought after effect. For
rare gas pressures on the order of 100 atmospheres, we estimate that a change
in the decay rate of order of 3\% could be observed. To increase the effect it
is necessary to find radiator atoms whose first excited state is radiatively
coupled to the ground state and dielectric atoms whose lowest excited state is
about 0.2eV above the energy of the excited state of the radiator. In this
limit, quenching will be negligible, but the detuning $\Delta$ is decreased
from the alkali-rare gas system by a factor of 50. At the same time, it is
necessary to achieve a high pressure for the dielectric atoms. A possible
system would be Li radiators with a high density sodium dielectric; the energy
mismatch of Li and Na is about 0.25eV, giving a correction factor to the
lithium decay rate of $1.3\times10^{-21}N$, where $N$ is the sodium dielectric
density in units of atoms/cm$^{3}.$

\section{Acknowledgments}

This research is supported by the National Science Foundation under Grants No.
PHY-0244841 and the FOCUS Center Grant. PRB would like to thank Peter Milonni
for helpful comments and Georg Raithel for a discussion concerning high
pressure sodium lamps.

\section{Appendix}

In this appendix, we calculate explicitly $G_{1,1}(\mathbf{R,}\omega)$ given
in Eq. (\ref{propagator}). The other $G_{m,m^{\prime}}(\mathbf{R,}\omega)$ are
calculated in a similar fashion. To carry out the angular integrations, one
expands $e^{i\mathbf{k}\cdot\mathbf{R}}$ as%
\begin{equation}
e^{i\mathbf{k}\cdot\mathbf{R}}=4\pi\sum_{m=-l}^{l}i^{l}Y_{lm}^{\ast
}(\mathbf{\hat{k}})Y_{lm}(\mathbf{\hat{R}})j_{l}(kR),
\label{spherical expansion}%
\end{equation}
uses the fact that $(\epsilon_{\mathbf{k}}^{(\lambda)})_{1}^{\ast}%
(\epsilon_{\mathbf{k}}^{(\lambda)})_{1}=\frac{1}{2}(1+\cos^{2}\theta)$
$=\frac{\sqrt{4\pi}}{3}[2Y_{00}(\mathbf{\hat{k}})+\frac{1}{\sqrt{5}}%
Y_{20}(\mathbf{\hat{k}})]$, and the orthogonality of the spherical harmonics,
to obtain%
\begin{equation}
G_{1,1}(\mathbf{R,}\omega)=\frac{1}{\sqrt{\pi}\omega^{3}}\int_{0}^{t}d\tau
\int_{-\infty}^{\infty}d\omega_{k}\omega_{k}^{3}e^{-i(\omega_{k}-\omega)\tau
}[2Y_{00}(\mathbf{\hat{R}})j_{0}(kR)-\frac{1}{\sqrt{5}}Y_{20}(\mathbf{\hat{R}%
})j_{2}(kR)] \label{a1}%
\end{equation}
The spherical Bessel function can written in terms of spherical Hankel
functions as $j_{l}(kR)=\frac{1}{2}[h_{l}(kR)+h_{l}^{\ast}(kR)]$, transforming
Eq. (\ref{a1}) into%
\[
G_{1,1}(\mathbf{R,}\omega)=\frac{1}{2\sqrt{\pi}}\int_{0}^{t}d\tau\int
_{-\infty}^{\infty}d\omega_{k}e^{-i(\omega_{k}-\omega)\tau}\{2Y_{00}%
(\mathbf{\hat{R}})[h_{0}(kR)+h_{0}^{\ast}(kR)]-\frac{1}{\sqrt{5}}%
Y_{20}(\mathbf{\hat{R}})[h_{2}(kR)+h_{2}^{\ast}(kR)]\}.
\]
In the calculation we always make the \textit{Wigner-Weisskopf }approximation.
Differences between $\omega,\omega_{0}$ and $\omega_{k}$ are neglected except
they appear as exponential factors. In integrating over $\omega_{k},$ the
$h_{l}^{\ast}(kR)$ terms give a contribution proportional to $\delta
(R/c+\tau)$ while the $h_{l}(kR)$ terms give a contribution proportional to
$\delta(R/c-\tau)$. We retain only the $\delta(R/c-\tau)$ contributions since
they correspond to the retarded solution (outgoing spherical waves). As a
consequence, we find%
\begin{equation}
G_{1,1}(\mathbf{R,}\omega)=\sqrt{4\pi}h_{0}(k_{0}R)Y_{0,0}(\mathbf{\hat{R}%
})-\frac{1}{2}\sqrt{\frac{4\pi}{5}}h_{2}(k_{0}R)Y_{2,0}(\mathbf{\hat{R}})
\end{equation}

\bibliographystyle{aaai-named}
\bibliography{articles}

\begin{thebibliography}{99}
\bibitem {note}The contributions from other terms may involve factors such as
$h_{0}(k_{0}R_{1})h_{2}(k_{0}R_{21})h_{2}(k_{0}R_{2})$. Direct expansion of
$h_{2}(k_{0}R_{21})$ gives a divergent result for the decay rate, when the
contribution from $R_{1}=R_{2}$ is excluded. Instead, we expand $h_{0}%
(k_{0}R_{1})$ in terms of $R_{21}$ and $R_{2}$ and integrate over these two
variables. This procedure leads to the finite results given in the paper. The
fact that we get divergent results when directly expanding $h_{2}(k_{0}%
R_{21})$ in terms of $R_{1}$ and $R_{2}$ can be traced to the fact that one
finds different values for the integrals at $R_{1}=R_{2}$ depending on whether
the limit $R_{1}=R_{2}$ is approached from above or below. If the contribution
from $R_{1}=R_{2}$ is included, it leads to a divergence that exactly cancels
the divergence from the first integral and again leads to the same result
given in the text.
\end{thebibliography}

\end{document}